\documentclass[pre,twocolumn,aps,10pt,superscriptaddress,notitlepage,nofootinbib]{revtex4-1}
\usepackage{epsfig,bm,amsmath,amssymb,amsfonts,graphicx,color,calc,epstopdf,stmaryrd}

\newcommand{\beq}{\begin{equation}}
\newcommand{\eeq}{\end{equation}}
\newcommand{\ba}{\begin{eqnarray}}
\newcommand{\ea}{\end{eqnarray}}

\newcommand{\B}[1]{{\bm{#1}}}

\newcommand{\ud}{\mathrm{d}}

\begin{document}

\title{
Breakdown of Nonlinear Elasticity in Stress-Controlled Thermal Amorphous Solids}

\author{Vladimir Dailidonis}
\affiliation{Bogolyubov Institute for Theoretical Physics, 03680 Kiev, Ukraine}

\author{Valery Ilyin}
\affiliation{Department of Chemical Physics, the Weizmann Institute of Science, Rehovot 76100, Israel}

\author{Itamar Procaccia}
 \affiliation{Department of Chemical Physics, the Weizmann Institute of Science, Rehovot 76100, Israel}

\author{Carmel A.B.Z. Shor}
\affiliation{Department of Chemical Physics, the Weizmann Institute of Science, Rehovot 76100, Israel}

\begin{abstract}
In recent work it was clarified that amorphous solids under strain control do not possess
nonlinear elastic theory in the sense that the shear modulus exists but nonlinear moduli
exhibit sample to sample fluctuations that grow without bound with the system size. More
relevant however for experiments are the conditions of stress control. In the present
Communication we show that also under stress control the shear modulus exists but higher
order moduli show unbounded sample to sample fluctuation. The unavoidable consequence is that
the characterization of stress-strain curves in experiments should be done with a stress-dependent
shear modulus rather than with nonlinear expansions.
\end{abstract}

\maketitle

Amorphous solids are born as a result of the glass transition, by cooling sufficiently rapidly certain pure liquids or mixtures of liquids such as to avoid the equilibrium phase transition to a crystalline solid. In recent years the theoretical understanding of amorphous solids is advancing, showing that they have rather peculiar mechanical properties. In particular it was clarified recently that generically amorphous solids do not have
a regular non-linear elasticity theory \cite{HentschelKarmakar11,DubeyProcaccia16,PRSS16}; describing the stress-strain relation with as a power series in the strain for example is untenable. This was shown first
 for athermal amorphous solids independently of the deformation protocol, and later for thermal amorphous
 solids under strain control conditions. For a wide range of (low) temperatures it is impossible to define non-linear elastic coefficients, as they do not converge to a specific value with the increase of the system size \cite{PRSS16}. The physical reason for this lack of convergence is the anomalous statistics of stress fluctuations in amorphous solids. It was also shown that in strain control conditions nonlinear expansion of stress in powers of strain are not necessary, since the stress-strain curves can be fully characterized using a theoretically meaningful (local) $\gamma$-dependent shear modulus $\mu(\gamma)$ \cite{DubeyProcaccia16}.

 Additional interest in this phenomenon has emerged due to its relation to the putative Gardner transition \cite{Ga85}
 which is believed to exist in generic glass formers \cite{12KPZF,13KPZF,fullRSB}. One of the main consequences of the Gardner
 transition, if it exists, is precisely the breakdown of nonlinear elasticity in the sense discussed above \cite{16BU}.
 This increases the motivation to identify the breakdown of nonlinear elasticity in experimental system.
 Experiments are however rarely done in strain controlled conditions since it is much easier to control the stress on a piece of material than its strain. It is therefore necessary to examine the elasticity theory of
 amorphous solids in stress control condition. In this Communication we examine the nonlinear elasticity theory of stress controlled simulations, where the strain is the fluctuating quantity. The central conclusion is that
 also in stress-controlled condition the nonlinear elasticity theory breaks down, and should not be used in the context of amorphous solids. Using a stress-controlled simulation takes us one step closer to systems used in experiments, and perhaps similar results can be obtained in future experimental measurements.

Consider the standard non-linear elasticity theory for a solid under simple shear strain, with $\gamma = \gamma_{xy}$ being the only non-zero component of the strain tensor. The stress is computed as a Taylor expansion around zero strain~\cite{LandauElasticity}:
\begin{equation}
 \sigma(\gamma) = B_1\gamma + \frac{1}{2!}B_2\gamma^2 + \frac{1}{3!}B_3\gamma^3 + \dots
 \label{sigmaPow}
\end{equation}
where $\sigma = \sigma_{xy}$ is the only non-zero component of the stress tensor and
\begin{equation}
 B_n \equiv \left.\frac{d^n\sigma}{d\gamma^n}\right|_{\gamma=0}.
\end{equation}
$B_1$ is the usual shear modulus that is usually denoted as $\mu$, $\mu\equiv B_1$.  Inverting the power series ~(\ref{sigmaPow})  will give an equivalent expression:
\begin{eqnarray}
\gamma(\sigma)&=&\gamma_0 +s_1\sigma+\frac{1}{2!}s_2\sigma^2+\frac{1}{3!}s_3\sigma^3\ldots\nonumber\\
&=&\gamma_0 +\frac{\partial \gamma(\sigma)}{\partial \sigma}\Big|_{\sigma=0}\sigma+\frac{1}{2!}\frac{\partial^2 \gamma(\sigma)}{\partial \sigma^2}\Big|_{\sigma=0}\sigma^2+\ldots
\label{gammaPow}
\end{eqnarray}
The relations between the coefficients in Eq.~(\ref{sigmaPow}) and Eq.~(\ref{gammaPow}) are given by \cite{Abramowitz}
\begin{eqnarray}
B_1&=&\frac{1}{s_1}\nonumber\\
B_2&=&-\frac{s_2}{s_1^3}\nonumber\\
B_3&=&\frac{3s_2^2-s_1 s_3}{s_1^5}\nonumber\\
B_4&=&\frac{10s_1 s_2 s_3-s_1^2s_4-15s_2^3}{s_1^7} \cdots \nonumber
\label{coef}
\end{eqnarray}

To compute the derivatives appearing in Eq.~(\ref{gammaPow}) in terms of strain fluctuations we need
a statistical-mechanical expression for the mean strain. Consider a system with $N$ particles in positions $\B R\equiv \{\B r_i\}_{i=1}^N$ in a volume $V$. When the external stress $\sigma^{\rm ext}$ is fixed, the ensemble average is provided by (see, e.g., \cite{DIMP14})
\begin{equation}
\gamma(\sigma^{\rm ext})=\frac{\int\gamma  e^{(-U(\gamma,{\bf R}) +V\sigma^{ext}\gamma)/T}\ud \gamma\ud {\bf R}}{\int e^{(-U(\gamma,{\bf R}) +V\sigma^{\rm ext}\gamma)/T}\ud \gamma\ud {\bf R}},
\label{Msear}
\end{equation}
where $T$ is the temperature in units where the Boltzmann constant is unity, $U(\gamma,{\bf R})$ is the energy of the system.

The computation of the derivatives in Eq.~(\ref{gammaPow}) is now straightforward \cite{RT10}, yielding
\begin{eqnarray}
s_1&=&\frac{V}{T}(\langle\gamma^2\rangle_0-\langle\gamma\rangle_0^2)\nonumber\\
s_2&=&\bigg(\frac{V}{T}\bigg)^2\big(\langle \gamma^3\rangle_0-3\langle\gamma\rangle_0\langle\gamma^2\rangle_0+2\langle\gamma\rangle_0^3\big)\nonumber\\
s_3&=&\bigg(\frac{V}{T}\bigg)^3\big(\langle \gamma^4\rangle_0-4\langle\gamma^3\rangle_0\langle\gamma\rangle_0-3\langle\gamma^2\rangle_0^2\nonumber\\
&+&12\langle\gamma^2\rangle_0\langle\gamma\rangle_0^2-6\langle\gamma\rangle_0^4\big)\nonumber\\
\ldots
\label{Ccum}
\end{eqnarray}
where the moments of the strain distribution at zero external shear stress are defined by
\begin{equation}
\langle\gamma^n\rangle_0=\frac{\int\gamma^n e^{-U(\gamma,{\bf R})/T} \ud \gamma\ud {\bf R}}{\int e^{-U(\gamma,{\bf R})/T}\ud \gamma\ud {\bf R}}.
\label{mom}
\end{equation}

The main question now is what is the statistics of strain fluctuation that determine these moments. In an amorphous solid the system is confined to a compact set in a \emph{restricted} domain of the configuration space; accordingly for each realization of the amorphous solid the integral (\ref{mom}) is computed over this set of configurations which are visited by the glass particles, confined around a given amorphous structure ~\cite{corrado2,DubeyProcaccia16}. This fact is the origin of sample-to-sample fluctuations;  these fluctuations may either decrease or increase with the system size, depending on the moment involved. This is precisely what
we need to learn using the numerical simulations described next.

Having the expressions (\ref{Ccum}) for the coefficients $s_n$ we will measure them below in numerical
simulations for different realizations of the same glass. We will then use Eqs.~(\ref{coef}) to determine
the moduli $B_n$. Needless to say, for each realization of the glass with $N$ particles we will
find a number for $B_n$. As explained, the interesting question is the sample-to-sample fluctuation, do they
decrease or increase with the system size. In other words, we will measure the distributions
of these objects and the moments of sample-to-sample variances
\begin{equation}
\overline{(\delta B_n)^2} \equiv \overline{(B_n-\overline{B_n})^2}\ , \label{var}
\end{equation}
where $\overline{(\bullet)}$ denotes the average over different realizations.

To simulate a system in which the strain is free to fluctuate we employ the ``variable shape Monte-Carlo''  technique, cf. Ref. \cite{PR81,PR80}, explained in more details in Ref. \cite{DIMP14}.
This method starts by defining a square box of {\em unit area} where the particles are at positions $\hat{ \B r}_i$. Next one defines a linear transformation $\B h$, taking the particles to positions $\B r_i$ via $\B r_i={\bf h}\cdot \hat{ \B r}_i$.  The actual area of the system becomes the determinant
$V=\mid {\bf h}\mid$. The Monte Carlo procedure employs two kinds of moves. Firstly one performs $n$ standard Monte Carlo moves
\begin{equation}
\hat{ \B r}^{new}_i=\hat{ \B r}^{old}_i+\delta \hat{ \B r},\hspace{4 mm} 1\le i\le N.
\label{Rmove}
\end{equation}
In this equation the $\alpha$ component of the displacement vector of a particle is given by
\begin{equation}
\delta \hat{  r}^{\alpha}=\Delta \hat{ r}_{max}(2\xi^\alpha-1),
\label{ParDisp}
\end{equation}
where $\Delta \hat{ \B r}_{max}$ is the maximum displacement and
$\xi^\alpha$ is an independent random number uniformly distributed between 0 and 1.
After these standard sweeps the transformation ${\bf h}$ changes according to
\begin{equation}
\B h^{new}=\B h^{old}+\delta\B h,
\label{hnew}
\end{equation}
where elements of the random symmetric matrix $\delta\B h$ are defined by
\begin{equation}
\delta h_{ij}=\Delta h_{max}(2\xi_{ij}-1),\hspace{4mm} i\le j.
\label{transH}
\end{equation}
Here $\Delta h_{max}$ is the maximum allowed change of a matrix element and
$\xi_{ij}$ is an independent random number uniformly distributed between 0 and 1.
\begin{figure}
\includegraphics[width=0.34\textwidth]{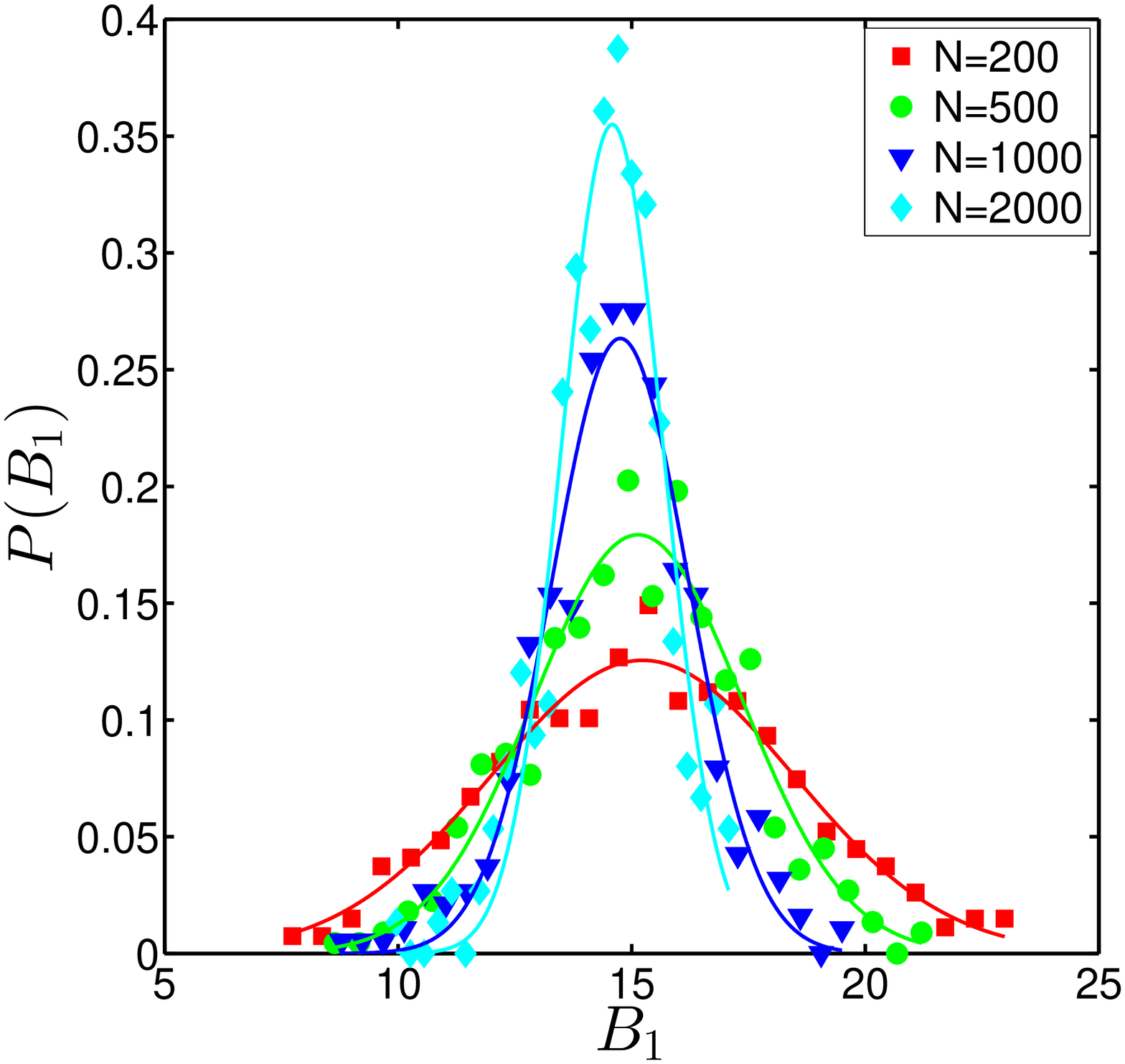}
\includegraphics[width=0.34\textwidth]{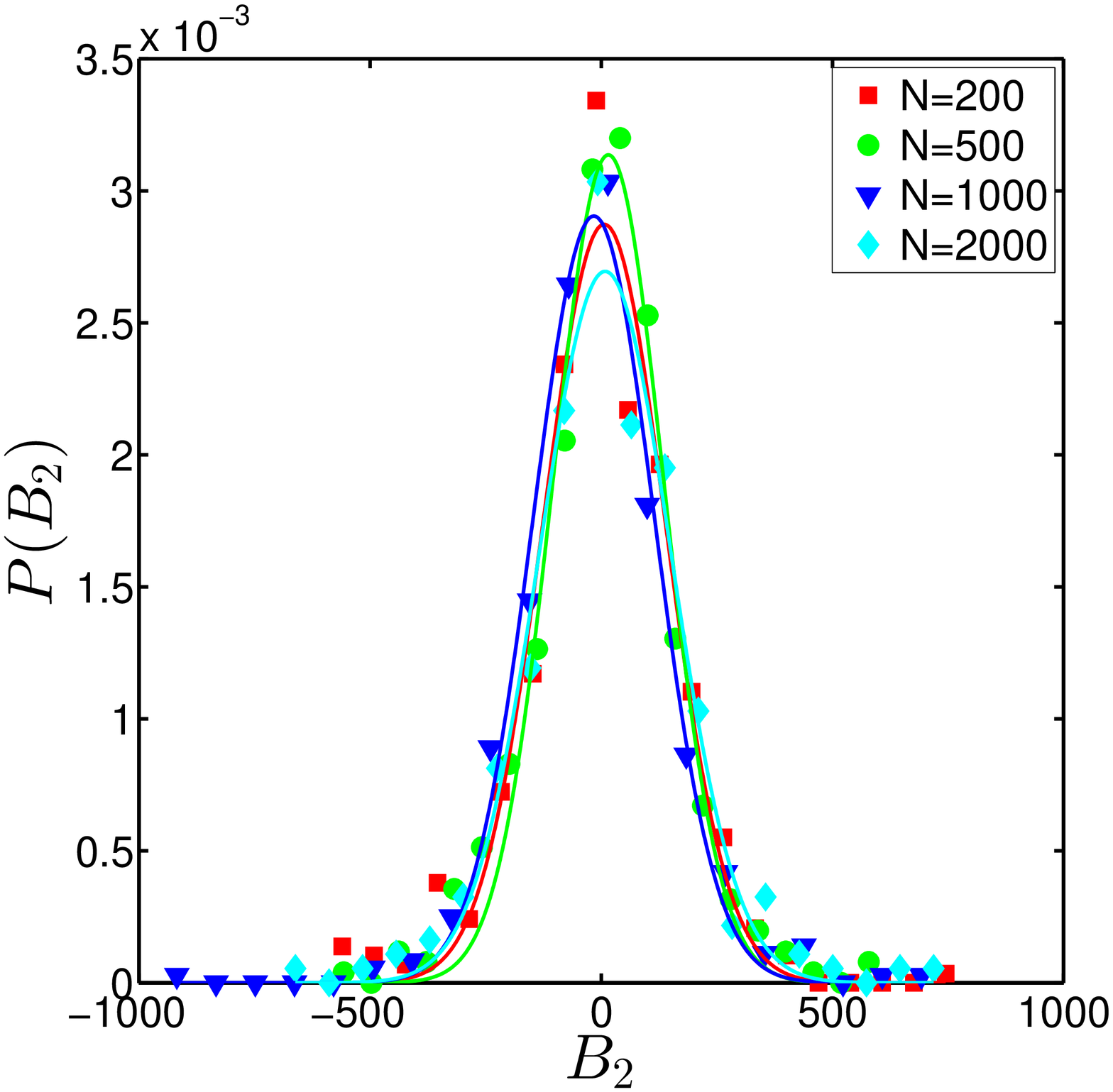}
\includegraphics[width=0.34\textwidth]{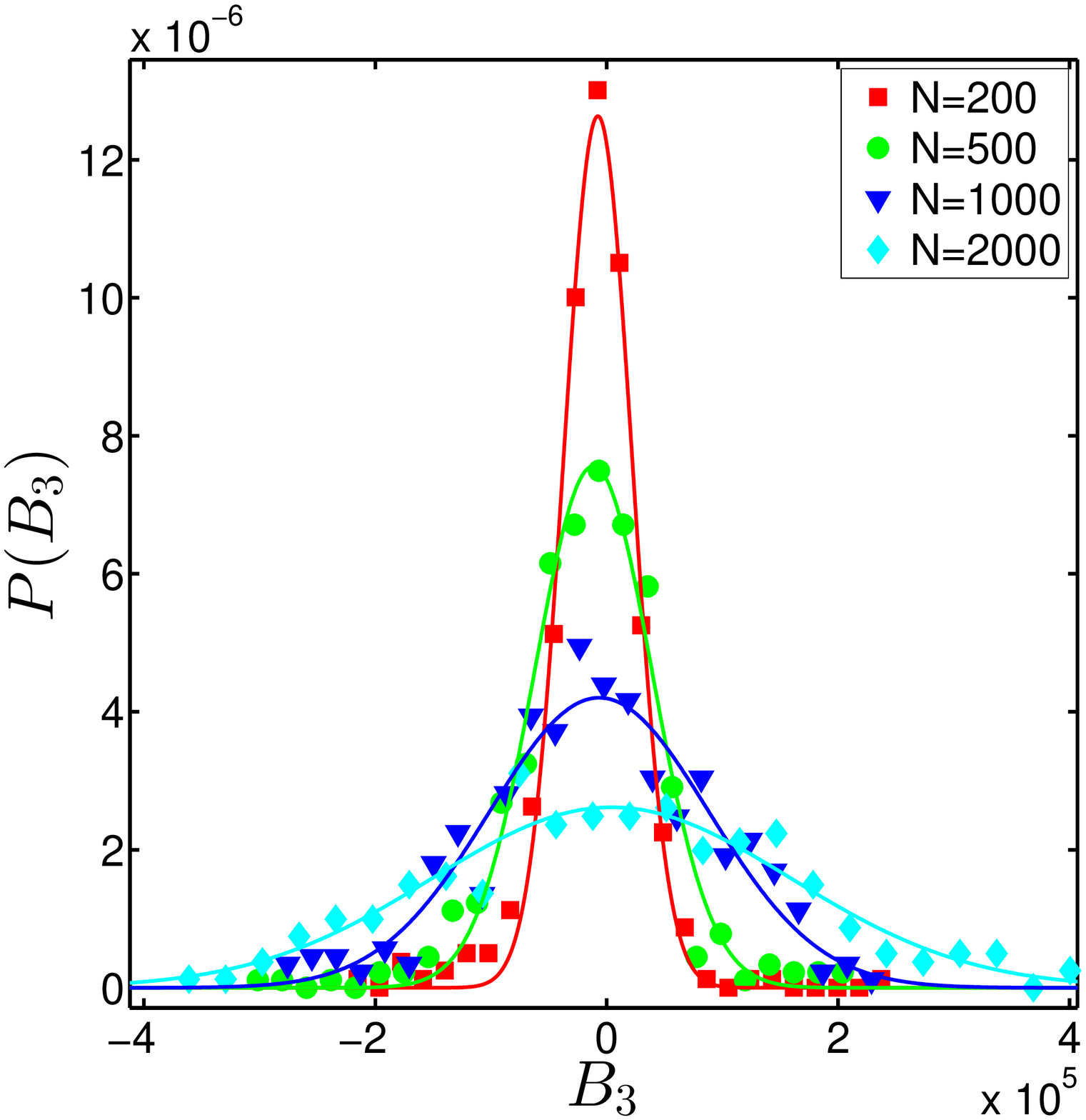}
 \caption{The distribution of values of the first three coefficients in the expansion Eq~(\ref{sigmaPow}) over
 the realizations, for different systems size. The temperature here is $T=0.05$.}
 \label{disFig}
 \end{figure}
In this particular case the  ${\bf h}$ matrix was chosen to induce simple shear conditions
\begin{equation}
{\bf h}=L\left(
\begin{array}{c c}
1&\gamma \\
0&1
\end{array}
\right)
=L\left(
\begin{array}{c c}
1&\delta h\\
0&1
\end{array}
\right),
\label{hG}
\end{equation}
where $L$ is the length of the square simulation box and $\gamma$ is the simple shear strain, with the volume of the system V=$L^2$ being conserved.
The value of $\Delta h_{max}$  and the maximum displacement of particle positions $\Delta \hat{ r}_{max}$ are selected to obtain a desired acceptance rate of $30\%$.
For each kind of move the trial configuration is accepted with probability
\begin{equation}
P_{tr}=\min \bigg[1,\exp\bigg(-\frac{\Delta G}{T}\bigg)\bigg].
\label{trial}
\end{equation}
Here $\Delta G$ is the {\em enthalpy change} due to the move. Since in this case we chose volume conserving transformations without any external stress was applied the expression is simply
\begin{equation}
\Delta G=U(\B \gamma+\B {\delta \gamma},\B r_i^{new})-
U(\B \gamma,\B r_i^{old})\ ,
\label{entG}
\end{equation}
where $\gamma$ is the system's strain.
In the current study no external stress is implemented, and thus the energy difference is only due to the affine deformation itself. Thus the strain $\gamma$ is free to fluctuate around some mean value. The reader should
note that for any given realization of the glass the mean value of $\gamma$ is not necessarily zero. Only
when these means are averaged over many glassy realization in the sense of the overline average Eq.(\ref{var}) the result should vanish.
  \begin{figure}
\includegraphics[width=0.34\textwidth]{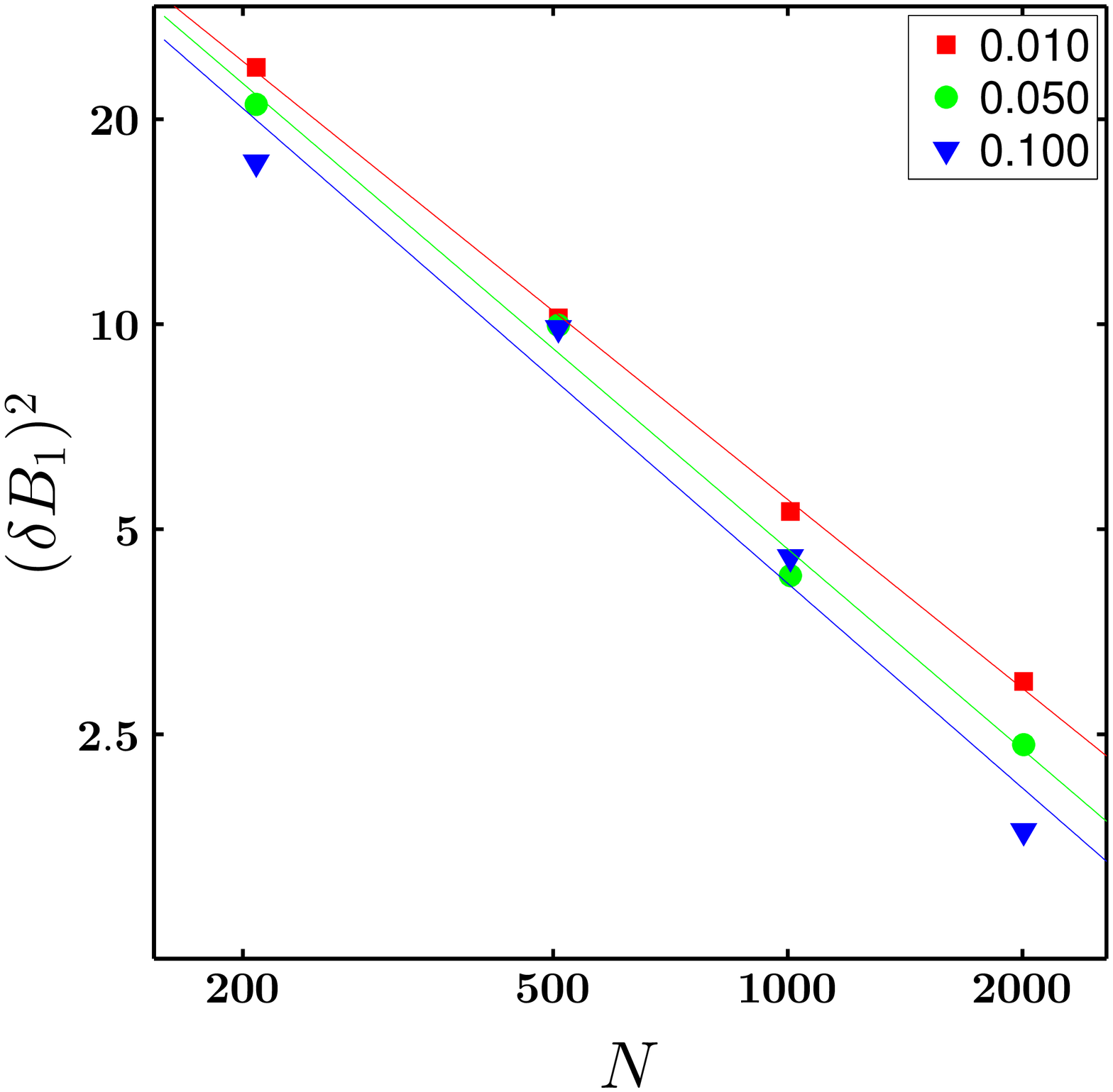}
\includegraphics[width=0.34\textwidth]{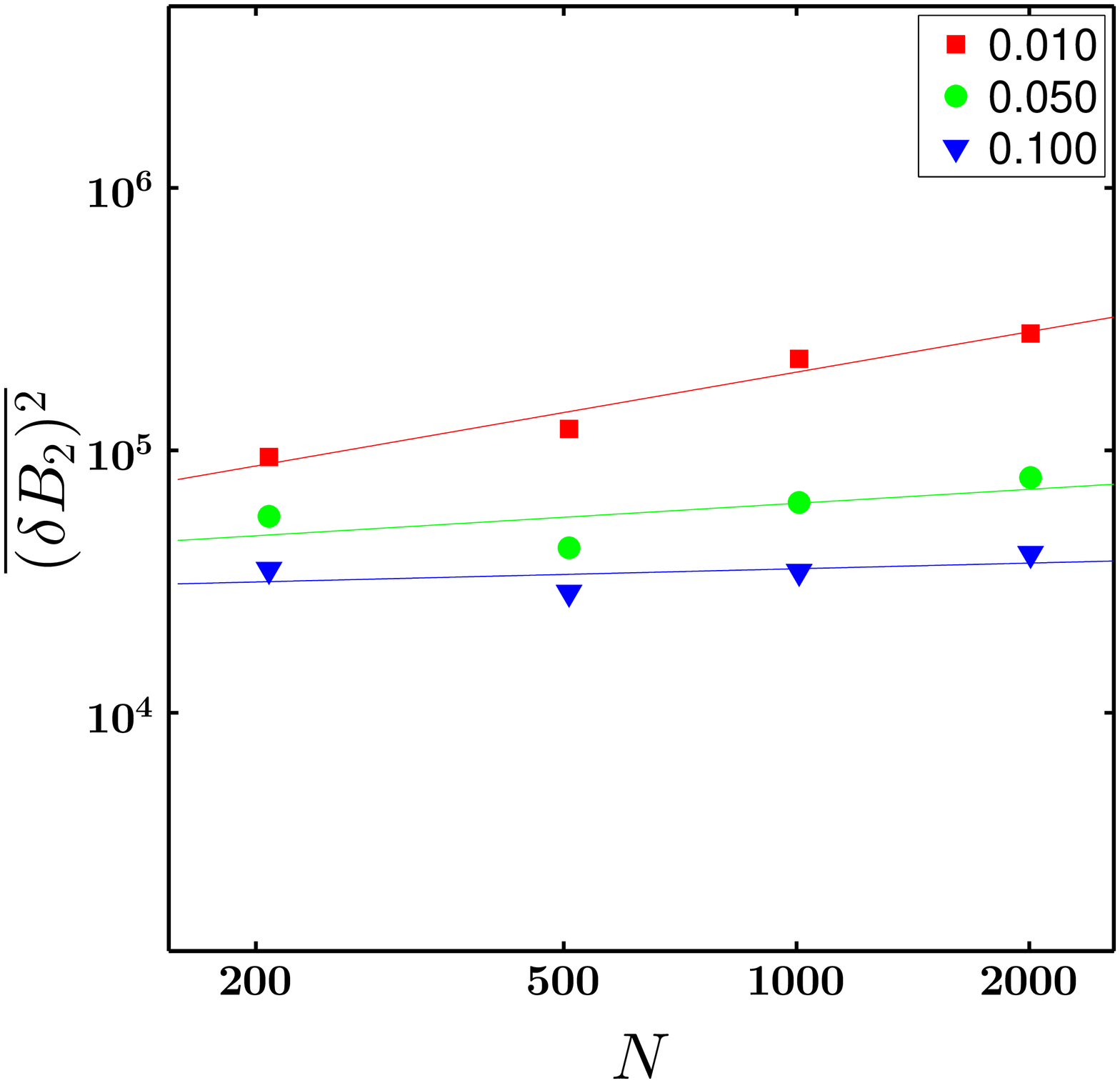}
\includegraphics[width=0.34\textwidth]{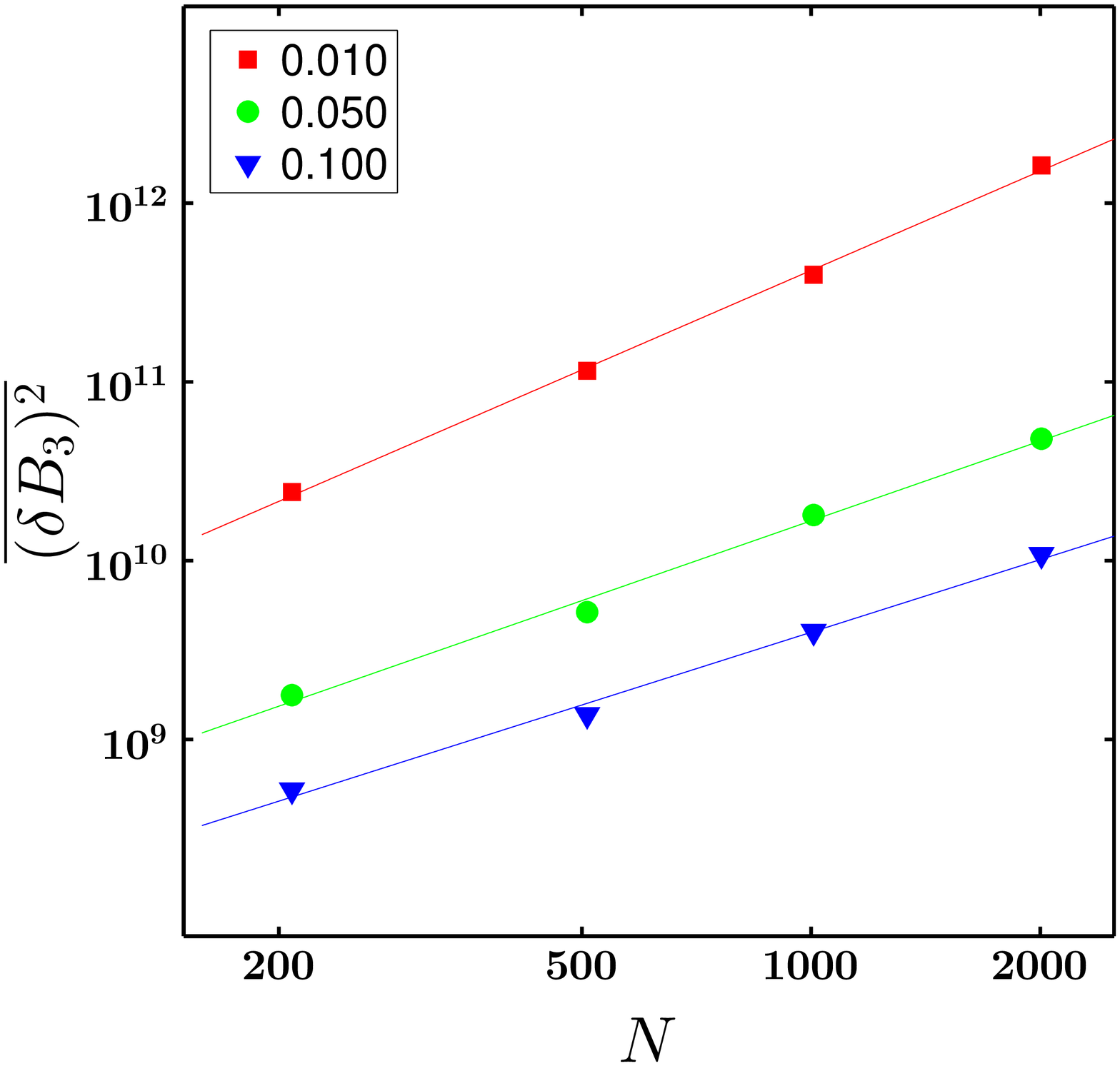}
 \caption{The system size dependence of the variances of the first three coefficients in the expansion Eq~(\ref{sigmaPow}). Here the temperatures are $T$=0.01, 0.05 and 0.1.}
 \label{Nfig}
 \end{figure}

To collect data we used the same glass former used in Ref.~\cite{PRSS16}, i.e.  a  two-dimensional Kob-Andersen 65:35  binary Lennard-Jones mixture \cite{94KA, KA2D}, slowly quenched form a liquid at $T=0.4$ down to $T=10^{-6}$ using molecular dynamics with quench rate $\dot{T} = 10^{-6}$. Here temperature and all other
quantities are measured in dimensionless units. The number density is $\rho=1.162$. Each sample was than relaxed in the target temperature for $i=5 \times 10^5$ MC swaps, in which the parameters   $\Delta h_{max}, \Delta \hat{  r}_{max}$ were calibrated. Than each realization was ran  with zero external stress $\sigma^{ext}=0$ for $\tau=5 \times 10^6$ MC swaps, from which the moments of the strain fluctuations $\langle \gamma^n \rangle$  were calculated. The number of steps $\tau$ was chosen to be sufficiently large to reach saturation of the measured quantities and independence of the averaging window. To calculate the variances across the ensembles, for each system size $N=\{ 200,500,1000,2000\}$  between 300 to 800 different realizations were prepared and allowed to fluctuate independently. In order to keep the data clean, realizations that were detected exhibiting fluctuations of the strain around two (or more) mean values were eliminated from the ensemble and were not considered for the calculation of the sample-to-sample variances.

In Fig.~\ref{disFig} we show the distributions of $B_1,B_2$, and $B_3$ over the different sets of realizations as explained above. The data presented here are for $T=0.050$, similar plots were made for $T=0.010, 0.100$. For each of the coefficients, and for each system size, the values of $B_n$  across the whole ensemble ($N=300-800$ realizations) were binned and normalized to produce the PDF.  As in previous reports (\cite{HentschelKarmakar11,PRSS16}), one can see that the distribution of the shear modulus $\mu=B_1$  converge to a delta-function at a temperature dependent sharp value $B_1$=15.9, 14.9, 14.1 at temperatures $T$=0.01, 0.05,
0.1 respectively. The second modulus $B_2$ does
not converge with the system size, and the variance of the third modulus $B_3$ diverges with the systems size.
It is therefore interesting to examine the variance of the distributions as a function of system size, as
 shown in Fig.~\ref{Nfig}.  The ensemble variances $\overline{(\delta B_k)^2}$ is presented in a log-log
 plot for different system sizes, and fitted to a linear curve. From this data we can extract the following
 scaling laws:
\begin{equation}
\overline{(\delta B_1)^2} \sim N^{\alpha_1} \ ,  \overline{(\delta B_2)^2} \sim N^{\alpha_2} \ , \overline{(\delta B_3)^2} \sim N^{\alpha_3} \ ,
\end{equation}

To determine the exponents we estimate the variance by fitting a Gaussian to the distribution. In our experience
this method is superior to measuring the variance from the raw data that may have outliers that reduce the
accuracy of the estimate.  We measure $\alpha_1 =-0.95\pm0.04$ with only slight temperature dependence.
For $\alpha_3$ we measure the values $\alpha_3=1.6\pm 0.24$.
The distribution of $B_2$ seems marginal, and indeed the measured value of the exponent $\alpha_2$ is close
to zero, except at the lowest temperature where there is an apparent weak divergence.
 This is in line with the results in the athermal case \cite{HentschelKarmakar11}, but differs from the thermal strain-controlled ensemble \cite{PRSS16}, where a divergence was found also for variance of $B_2$.

It is interesting to relate these findings to a recent theoretical work~\cite{16BU} predicting a so-called Gardner transition~\cite{Ga85} in
thermal glass forming liquids \cite{fullRSB,16BU}. The phenomenon seen in this transition is that at some temperature, lower than the glass transition temperature, there appears a qualitative change in the free-energy
landscape, generating a rough surface with arbitrarily small barriers between local minima. The connection to the
present work is that this is accompanied by a breakdown of nonlinear elasticity in much the same
way reported here. The available theory pertains to a mean field treatment and comparison of exponents is
probably not warranted. Nevertheless it is interesting that the shear modulus is expected to exist, and the
variances of $B_k$ with $k\ge 3$ are expected to diverge with the system size, in agreement with the predictions
of Ref.~\cite{HentschelKarmakar11} and the findings of the present paper.  In Ref.~\cite{16BU} it is also predicted that the phenomenon should disappear when the system is heated above the (protocol dependent) Gardner temperature. Whether this is occurring before the shear modulus itself vanishes or not depends on the
speed of quench from high to low temperatures. A search of a Gardner temperature
would require repeating our analysis on extremely slowly quenched glasses as a way to provide a good separation
of the Gardner point and the point of disappearance of the shear modulus~\cite{16BU}. Such an analysis is beyond the scope of the present Communication but appears to be a worthwhile endeavor for future research.

Finally we should discuss the implications of our finding to stress-controlled experiments. Clearly, one can
repeat the procedure described above at a given stress value instead of zero stress. For any given stress
one can determine the mean strain, plot a stress vs. strain curve and determined the local slope.
The prediction that is implied in the present Communication is that this local slope will be
equivalent to the value of the shear modulus $\mu=B_1$ that can be measured from the moments of the strain fluctuations. There is no point in attempting to fit a nonlinear stress vs. strain curve since
the nonlinear moduli have no theoretical value. This prediction warrants a careful experimental
verification.

\acknowledgments

This work has been supported in part  by the Minerva foundation with funding from the Federal German Ministry for Education and Research, and by the Israel Science Foundation (Israel Singapore Program).

\bibliographystyle{mioaps}
\bibliography{LJ}

\begin{thebibliography}{17}
\expandafter\ifx\csname natexlab\endcsname\relax\def\natexlab#1{#1}\fi
\expandafter\ifx\csname bibnamefont\endcsname\relax
  \def\bibnamefont#1{#1}\fi
\expandafter\ifx\csname bibfnamefont\endcsname\relax
  \def\bibfnamefont#1{#1}\fi
\expandafter\ifx\csname citenamefont\endcsname\relax
  \def\citenamefont#1{#1}\fi
\expandafter\ifx\csname url\endcsname\relax
  \def\url#1{\texttt{#1}}\fi
\expandafter\ifx\csname urlprefix\endcsname\relax\def\urlprefix{URL }\fi
\providecommand{\bibinfo}[2]{#2}
\providecommand{\eprint}[2][]{\url{#2}}

\bibitem[{\citenamefont{Hentschel et~al.}(2011)\citenamefont{Hentschel,
  Karmakar, Lerner, and Procaccia}}]{HentschelKarmakar11}
\bibinfo{author}{\bibfnamefont{H.~G.~E.} \bibnamefont{Hentschel}},
  \bibinfo{author}{\bibfnamefont{S.}~\bibnamefont{Karmakar}},
  \bibinfo{author}{\bibfnamefont{E.}~\bibnamefont{Lerner}}, \bibnamefont{and}
  \bibinfo{author}{\bibfnamefont{I.}~\bibnamefont{Procaccia}},
  \bibinfo{journal}{Phys. Rev. E} \textbf{\bibinfo{volume}{83}},
  \bibinfo{pages}{061101} (\bibinfo{year}{2011}).

\bibitem[{\citenamefont{Dubey et~al.}(2016)\citenamefont{Dubey, Procaccia,
  Shor, and Singh}}]{DubeyProcaccia16}
\bibinfo{author}{\bibfnamefont{A.~K.} \bibnamefont{Dubey}},
  \bibinfo{author}{\bibfnamefont{I.}~\bibnamefont{Procaccia}},
  \bibinfo{author}{\bibfnamefont{C.~A. B.~Z.} \bibnamefont{Shor}},
  \bibnamefont{and} \bibinfo{author}{\bibfnamefont{M.}~\bibnamefont{Singh}},
  \bibinfo{journal}{Phys. Rev. Lett.} \textbf{\bibinfo{volume}{116}},
  \bibinfo{pages}{085502} (\bibinfo{year}{2016}).

\bibitem[{\citenamefont{Procaccia et~al.}(2016)\citenamefont{Procaccia,
  Rainone, Shor, and Singh}}]{PRSS16}
\bibinfo{author}{\bibfnamefont{I.}~\bibnamefont{Procaccia}},
  \bibinfo{author}{\bibfnamefont{C.}~\bibnamefont{Rainone}},
  \bibinfo{author}{\bibfnamefont{C.~A. B.~Z.} \bibnamefont{Shor}},
  \bibnamefont{and} \bibinfo{author}{\bibfnamefont{M.}~\bibnamefont{Singh}},
  \bibinfo{journal}{Phys. Rev. E} \textbf{\bibinfo{volume}{93}},
  \bibinfo{pages}{063003} (\bibinfo{year}{2016}).

\bibitem[{\citenamefont{Gardner}(1985)}]{Ga85}
\bibinfo{author}{\bibfnamefont{E.}~\bibnamefont{Gardner}},
  \bibinfo{journal}{Nuclear Physics B} \textbf{\bibinfo{volume}{257}},
  \bibinfo{pages}{747} (\bibinfo{year}{1985}).

\bibitem[{\citenamefont{Kurchan et~al.}(2012)\citenamefont{Kurchan, Parisi, and
  Zamponi}}]{12KPZF}
\bibinfo{author}{\bibfnamefont{J.}~\bibnamefont{Kurchan}},
  \bibinfo{author}{\bibfnamefont{G.}~\bibnamefont{Parisi}}, \bibnamefont{and}
  \bibinfo{author}{\bibfnamefont{F.}~\bibnamefont{Zamponi}},
  \bibinfo{journal}{Journal of Statistical Mechanics: Theory and Experiment}
  \textbf{\bibinfo{volume}{2012}}, \bibinfo{pages}{P10012}
  (\bibinfo{year}{2012}).

\bibitem[{\citenamefont{Kurchan et~al.}(2013)\citenamefont{Kurchan, Parisi,
  Urbani, and Zamponi}}]{13KPZF}
\bibinfo{author}{\bibfnamefont{J.}~\bibnamefont{Kurchan}},
  \bibinfo{author}{\bibfnamefont{G.}~\bibnamefont{Parisi}},
  \bibinfo{author}{\bibfnamefont{P.}~\bibnamefont{Urbani}}, \bibnamefont{and}
  \bibinfo{author}{\bibfnamefont{F.}~\bibnamefont{Zamponi}},
  \bibinfo{journal}{J.Phys. Chem. B} \textbf{\bibinfo{volume}{117}},
  \bibinfo{pages}{12979} (\bibinfo{year}{2013}).

\bibitem[{\citenamefont{Charbonneau et~al.}(2014)\citenamefont{Charbonneau,
  Kurchan, Parisi, Urbani, and Zamponi}}]{fullRSB}
\bibinfo{author}{\bibfnamefont{P.}~\bibnamefont{Charbonneau}},
  \bibinfo{author}{\bibfnamefont{J.}~\bibnamefont{Kurchan}},
  \bibinfo{author}{\bibfnamefont{G.}~\bibnamefont{Parisi}},
  \bibinfo{author}{\bibfnamefont{P.}~\bibnamefont{Urbani}}, \bibnamefont{and}
  \bibinfo{author}{\bibfnamefont{F.}~\bibnamefont{Zamponi}},
  \bibinfo{journal}{Nat. Comm.} \textbf{\bibinfo{volume}{5}},
  \bibinfo{pages}{3725} (\bibinfo{year}{2014}).

\bibitem[{\citenamefont{Biroli and Urbani}(2016)}]{16BU}
\bibinfo{author}{\bibfnamefont{G.}~\bibnamefont{Biroli}} \bibnamefont{and}
  \bibinfo{author}{\bibfnamefont{P.}~\bibnamefont{Urbani}},
  \bibinfo{journal}{Nature Physics} \textbf{\bibinfo{volume}{12}},
  \bibinfo{pages}{1130–1133} (\bibinfo{year}{2016}).

\bibitem[{\citenamefont{Landau and Lifshitz}(1959)}]{LandauElasticity}
\bibinfo{author}{\bibfnamefont{L.~D.} \bibnamefont{Landau}} \bibnamefont{and}
  \bibinfo{author}{\bibfnamefont{E.~M.} \bibnamefont{Lifshitz}},
  \emph{\bibinfo{title}{Course of Theoretical Physics Vol 7: Theory of
  Elasticity}} (\bibinfo{publisher}{Pergamon Press}, \bibinfo{year}{1959}).

\bibitem[{\citenamefont{Abramowitz and Stegun}(1965)}]{Abramowitz}
\bibinfo{author}{\bibfnamefont{M.}~\bibnamefont{Abramowitz}} \bibnamefont{and}
  \bibinfo{author}{\bibfnamefont{I.}~\bibnamefont{Stegun}},
  \emph{\bibinfo{title}{Handbook of Mathematical Functions}}
  (\bibinfo{publisher}{Dover Publications}, \bibinfo{year}{1965}).

\bibitem[{\citenamefont{Dailidonis et~al.}(2014)\citenamefont{Dailidonis,
  Ilyin, Mishra, and Procaccia}}]{DIMP14}
\bibinfo{author}{\bibfnamefont{V.}~\bibnamefont{Dailidonis}},
  \bibinfo{author}{\bibfnamefont{V.}~\bibnamefont{Ilyin}},
  \bibinfo{author}{\bibfnamefont{P.}~\bibnamefont{Mishra}}, \bibnamefont{and}
  \bibinfo{author}{\bibfnamefont{I.}~\bibnamefont{Procaccia}},
  \bibinfo{journal}{Phys. Rev. E} \textbf{\bibinfo{volume}{90}},
  \bibinfo{pages}{052402} (\bibinfo{year}{2014}).

\bibitem[{\citenamefont{Rodríguez and Tsallis}(2010)}]{RT10}
\bibinfo{author}{\bibfnamefont{A.}~\bibnamefont{Rodríguez}} \bibnamefont{and}
  \bibinfo{author}{\bibfnamefont{C.}~\bibnamefont{Tsallis}},
  \bibinfo{journal}{J. Math. Phys.} \textbf{\bibinfo{volume}{51}},
  \bibinfo{pages}{073301} (\bibinfo{year}{2010}).

\bibitem[{\citenamefont{Rainone et~al.}(2015)\citenamefont{Rainone, Urbani,
  Yoshino, and Zamponi}}]{corrado2}
\bibinfo{author}{\bibfnamefont{C.}~\bibnamefont{Rainone}},
  \bibinfo{author}{\bibfnamefont{P.}~\bibnamefont{Urbani}},
  \bibinfo{author}{\bibfnamefont{H.}~\bibnamefont{Yoshino}}, \bibnamefont{and}
  \bibinfo{author}{\bibfnamefont{F.}~\bibnamefont{Zamponi}},
  \bibinfo{journal}{Phys. Rev. Lett.} \textbf{\bibinfo{volume}{114}},
  \bibinfo{pages}{015701} (\bibinfo{year}{2015}).

\bibitem[{\citenamefont{Parrinello and Rahman}(1981)}]{PR81}
\bibinfo{author}{\bibfnamefont{M.}~\bibnamefont{Parrinello}} \bibnamefont{and}
  \bibinfo{author}{\bibfnamefont{A.}~\bibnamefont{Rahman}},
  \bibinfo{journal}{Journal of Applied Physics} \textbf{\bibinfo{volume}{52}},
  \bibinfo{pages}{7182} (\bibinfo{year}{1981}).

\bibitem[{\citenamefont{Parrinello and Rahman}(1980)}]{PR80}
\bibinfo{author}{\bibfnamefont{M.}~\bibnamefont{Parrinello}} \bibnamefont{and}
  \bibinfo{author}{\bibfnamefont{A.}~\bibnamefont{Rahman}},
  \bibinfo{journal}{Phys. Rev. Lett.} \textbf{\bibinfo{volume}{45}},
  \bibinfo{pages}{1196} (\bibinfo{year}{1980}).

\bibitem[{\citenamefont{Kob and Andersen}(1994)}]{94KA}
\bibinfo{author}{\bibfnamefont{W.}~\bibnamefont{Kob}} \bibnamefont{and}
  \bibinfo{author}{\bibfnamefont{H.~C.} \bibnamefont{Andersen}},
  \bibinfo{journal}{Physical review letters} \textbf{\bibinfo{volume}{73}},
  \bibinfo{pages}{1376} (\bibinfo{year}{1994}).

\bibitem[{\citenamefont{Brüning et~al.}(2009)\citenamefont{Brüning, St-Onge,
  Patterson, and Kob}}]{KA2D}
\bibinfo{author}{\bibfnamefont{R.}~\bibnamefont{Brüning}},
  \bibinfo{author}{\bibfnamefont{D.~A.} \bibnamefont{St-Onge}},
  \bibinfo{author}{\bibfnamefont{S.}~\bibnamefont{Patterson}},
  \bibnamefont{and} \bibinfo{author}{\bibfnamefont{W.}~\bibnamefont{Kob}},
  \bibinfo{journal}{Journal of Physics: Condensed Matter}
  \textbf{\bibinfo{volume}{21}}, \bibinfo{pages}{035117}
  (\bibinfo{year}{2009}).

\end{thebibliography}

\end{document}